\documentclass{iau}

\usepackage{amsmath}
\usepackage{graphicx}
\usepackage{multirow}
\usepackage{subfig}
\usepackage{hyperref}

\begin{document}

\lefttitle{Omkumar et al.}
\righttitle{IAU Symposium 379: SMC [Fe/H] estimates from Str\"{o}mgren photometry }

\jnlPage{1}{7}
\jnlDoiYr{2023}
\doival{10.1017/xxxxx}

\aopheadtitle{Proceedings of IAU Symposium 379}
\editors{P. Bonifacio,  M.-R. Cioni \& F. Hammer, eds.}

\title{Str\"{o}mgren photometric metallicity of the Small Magellanic Cloud stars using Gaia DR3-XP spectra}
\author{Abinaya O. Omkumar$^{1,2,3}$,
Smitha Subramanian$^{1}$, Maria-Rosa L. Cioni$^{2}$, Jos de Bruijne$^{4}$}
\affiliation{$^{1}$ Indian Institute of Astrophysics, Koramangala II Block, Bangalore-560034, India\\
$^{2}$Leibniz-Institut f\"ur Astrophysik Potsdam (AIP), An der Sternwarte 16. D-14482 Potsdam, Germany\\
$^{3}$Institut f\"{u}r Physik und Astronomie, Universit\"{a}t Potsdam, Haus 28, Karl-Liebknecht-Str. 24/25, D-14476 Golm (Potsdam), Germany\
$^{4}$European Space Agency, ESTEC, Noordwijk, Netherlands}

\begin{abstract}
Observational studies have identified several sub-structures in different regions of the Magellanic Clouds, the nearest pair of interacting dwarf satellites of the Milky Way. By studying the metallicity of the sources in these sub-structures, we aim to shed light on the possible origin of these sub-structures. Spectroscopic metallicities exist only for a few thousand sources, mostly giant stars located in specific regions of the galaxies. These metallicities come from different instruments at various spectral resolutions, and systematic uncertainties hamper comparisons and draw firm conclusions about their origin. The third data release of \textit{Gaia} has provided us with $\sim$ 0.17 million XP spectra of the different stellar populations in the SMC alone as faint as $\sim$ 18 mags in the G band, which are spread across $\sim$ 10$^\circ$ from the SMC centre. We aim to determine the metallicities of these sources based on synthetic Str\"{o}mgren photometry derived from XP spectra and produce a high-resolution metallicity map of the SMC. Our metallicity gradient estimate of the SMC turns out to be --0.062 $\pm$ 0.009 dex/deg. This is comparable with the previous estimates, which also validates our method of metallicity estimation.  We aim to apply this method to other stellar populations and to the LMC to create a high-resolution metallicity map of the Magellanic Clouds.
\end{abstract}

\begin{keywords}
galaxies: Magellanic Clouds, galaxies: abundances, galaxies: evolution
\end{keywords}

\maketitle

\section{Introduction}

The Large Magellanic Cloud (LMC) and the Small Magellanic Cloud (SMC) are both gas-rich interacting dwarf irregulars. Evidence shows that the LMC--SMC pair, also known as the Magellanic Clouds (MCs), interacts with the Milky Way. And their proximity ($\sim$ 55 kpc) makes them an ideal laboratory in the Local Group to study galaxy interaction and evolution processes in detail. Previous studies  (\citealt{2016MNRAS.459..239M,2017MNRAS.468.1349P,2018ApJ...858L..21M,2019MNRAS.482L...9B,2021MNRAS.505.2020E}) have identified multiple signatures of interactions in the form of stellar sub-structures, over-densities and gaseous structures in and around the MCs. Apart from these above-mentioned sub-structures, some dual populations of intermediate/old stars (red clump; \citealt{2021MNRAS.500.2757O} and red giant branch stars; \citealt{2014MNRAS.442.1680D}) have also been identified and studied. The results showed that they have different kinematics and are located at different distances. This could suggest that the dual populations formed during the interaction between the MCs and the stripped population's kinematics have been altered. If so, we expect similar metallicity ([Fe/H]) among the sources in these populations. Investigating the nature and origin of these sub-structures is essential for a comprehensive understanding of the consequences of dynamical interactions. But to do so, we need a metallicity estimate of a homogenous sample spread across the entire SMC, including its outskirts, where a plethora of sub-structures have been identified. Until now, we have spectroscopic metallicities of about a few thousand sources in the SMC from various instruments with different spectral resolutions. Comparing these estimates, which have different systematic uncertainties, is not trivial, and the results obtained will not be statistically significant. Hence, we need to use another standard indirect method to obtain the [Fe/H] of a larger sample. \\
\indent \cite{2021ApJ...909..150G} presented a machine-learning method to obtain photometric metallicity estimates for the Magellanic Cloud red giants from \textit{Gaia} Data Release 2 (DR2). Their predicted metallicity estimates of the MCs were comparable with the previous studies, but their method depends on their training sample. Also, the DR2 data was incomplete in the central regions of the MCs, and after applying all their quality filters, their final sample consisted of $\sim$ 36,000 SMC sources. In the latest \textit{Gaia} (DR3), $\sim$ 0.17 million XP spectra consisting of different stellar populations spread across a $\sim$ 10$^\circ$ of the SMC is provided as ancillary data. This is the largest spectroscopic dataset of the SMC obtained so far.  In this work, we used a subset of this dataset ($\sim$ 80,000 giant stars) to estimate metallicity using the synthetic Str\"{o}mgren photometry from the XP spectra. Str\"{o}mgren photometry has proved to be one of the reliable methods to determine the metallicity of giant and sub-giant stars \citep{2000A&A...355..994H}. Their calibration is based on a homogeneous sample of globular clusters ($\omega$ Centauri, M22 and M55), which are mostly metal-poor. When we apply this calibration relation to a sample consisting of both metal-poor giants and relatively metal-rich younger giants, the calibration would work poorly on the metal-rich giants, since such calibrators were absent \citep{2021A&A...656A.155L}. This is due to age-metallicity degeneracy. Although photometric metallicity estimates were proven to be a great tool for obtaining metallicity estimates, one must be careful to consider the age-metallicity degeneracy. In the case of our sample, it does not include many younger giants, so the effect of age-metallicity degeneracy is less likely to affect our estimates.

\section{Gaia sample selection}
\begin{figure}
    \centering
    \subfloat{\includegraphics[width=0.6\textwidth]{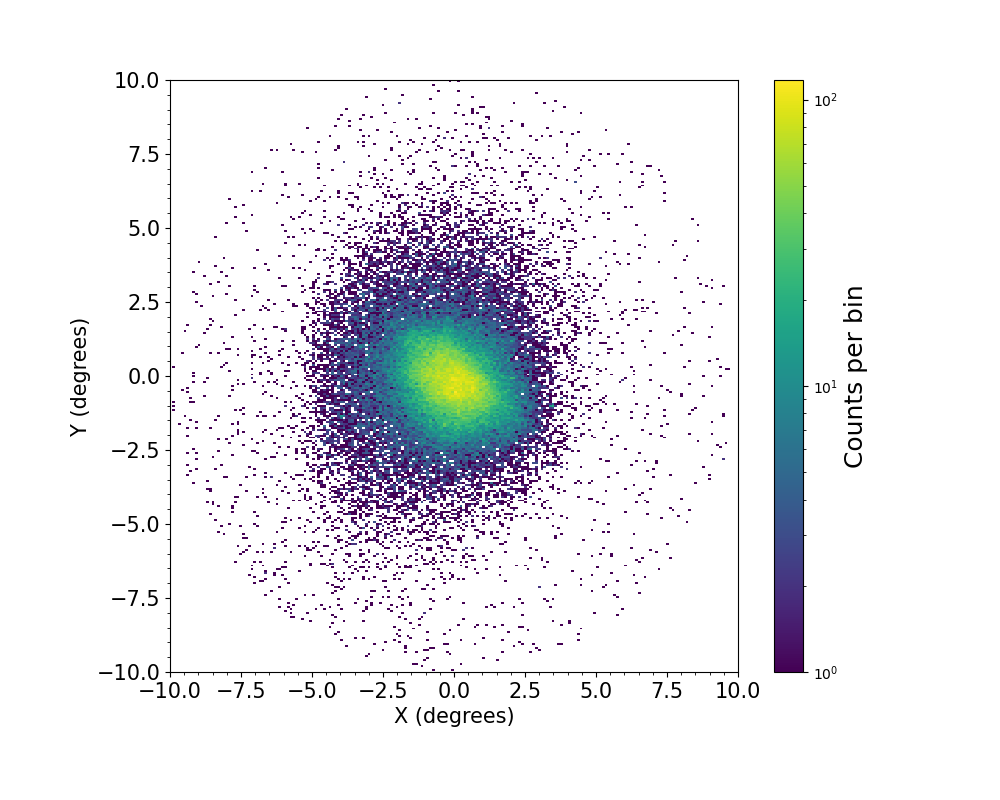}}
    
    \caption{Spatial distribution of the selected sources in the dereddened colour range 0.5 $<$ (b -- y)$_0$ $<$ 1.1 mag in XY plane centred on the SMC. The increasing number density is shown with the colour scale from blue to yellow.} 
    \label{fig:xy}
\end{figure}

We selected \textit{Gaia} DR3 sources within $\sim$ 10$^\circ$ of the SMC centre (RA = 12$\rlap{.}^{\circ}$80 and Dec = --73$\rlap{.}^{\circ}$15) with magnitudes G $<$ 20.5. This resulted in 4,710,809 sources, which further reduced to 171,060 sources having Gaia DR3 XP spectra. We also applied additional criteria on parallax and proper motions \citep{2021A&A...649A...7G, 2018A&A...616A..17A,  2021MNRAS.500.2757O} only to select probable SMC sources. Our final sample contained 151,530 sources. We used this source list to query the XP spectra from the Gaia archive (Datalink-products). Currently, we have estimated the [Fe/H] for only a subset ($\sim$ 80,000 sources) of the final sample in the dereddened colour range 0.5 $<$ (b -- y)$_0$ $<$ 1.1 mag for which the [Fe/H] calibrations were available in the literature. The spatial distribution of this subset of data is shown in Figure \ref{fig:xy}.

\section{Synthetic Photometry using GaiaXPy}
We derived Str\"{o}mgren magnitudes (v, b and y) using the GaiaXPy tool provided by the \textit{Gaia} consortium. This tool allows the generation of synthetic photometry in a set of desired systems from the input internally-calibrated continuously-represented mean spectra (see \cite{2022arXiv220606205M} for more details). We then corrected for the interstellar extinction by translating the extinction coefficient in the visual band A$_V$ into A$_v$, A$_b$ and A$_y$, where A$_V$=A$_0$/1.003 \citep{2014MNRAS.443.2907S} and A$_0$ is the extinction at 547.7 nm. Using these extinction-corrected magnitudes, we proceeded to estimate the [Fe/H]. 

\section{Estimation of [Fe/H]}
\begin{figure}
    \centering
    \subfloat{\includegraphics[width=0.5\columnwidth]{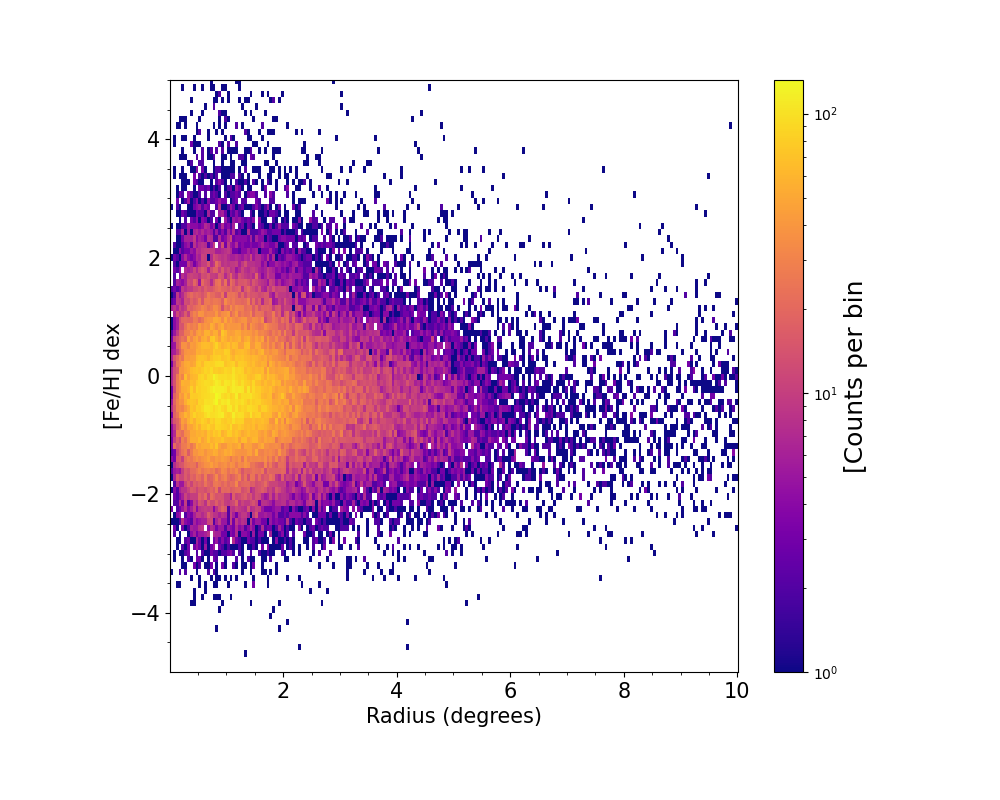}}
    \subfloat{\includegraphics[width=0.47\columnwidth]{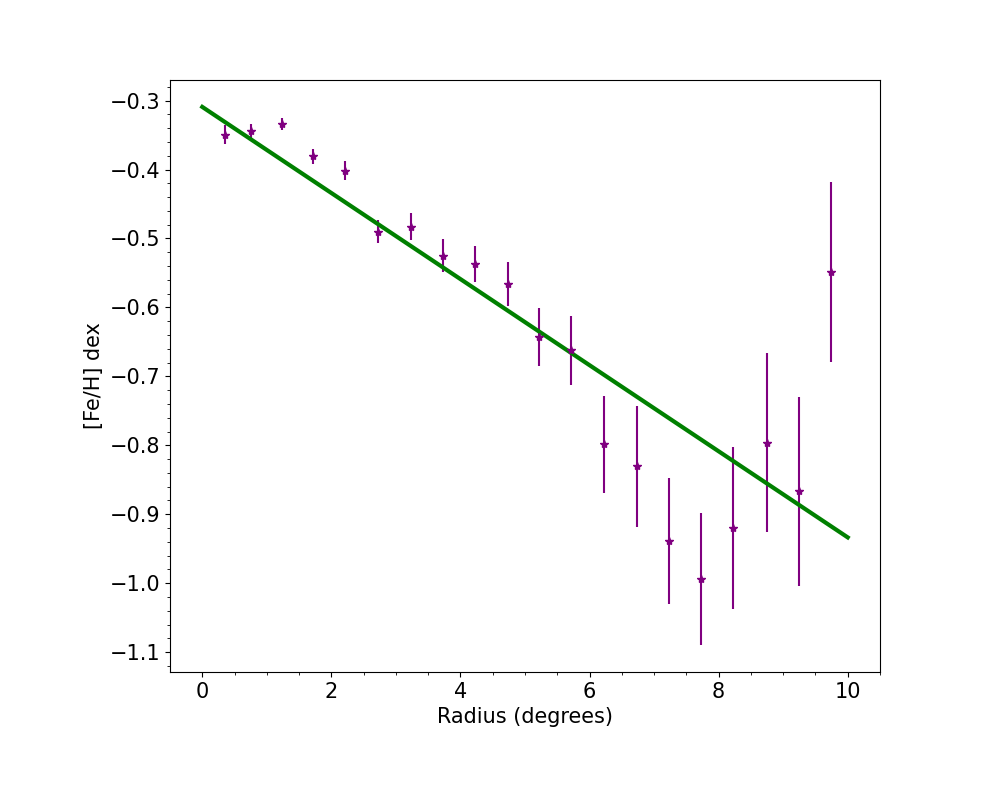}}  
    \caption{\textit{Left Panel}: Estimated [Fe/H] of the subset of selected (red in Figure \ref{fig:xy}) Gaia DR3 SMC sources as a function of radius.  \textit{Right Panel}: Median metallicities for the sources in the left panel as a function of radius. The linear fit corresponds to a gradient of --0.062 $\pm$ 0.009 dex/deg.} 
    \label{fig:feh}
\end{figure}

\cite{2000A&A...355..994H} provided the calibration equation to calculate the iron abundance [Fe/H] that is applicable for sources in the dereddened colour range 0.5 $<$ (b -- y)$_0$ $<$ 1.1 mag. This is reproduced here in equation \ref{equation:feh}. We used our extinction-corrected colours in this equation and estimated the [Fe/H]  of each of the sources. We plotted a 2D Hess diagram of the estimated [Fe/H] (dex) as a function of radius (degrees) from the SMC centre in the left panel of Figure \ref{fig:feh}. The increasing stellar density is indicated by the colour from blue to yellow. From the figure, it is clear that we do see a larger spread in the [Fe/H] values, but the stellar density indicates that very few sources have very low and high [Fe/H] values. To better understand these values, we binned the dataset in concentric shells of 0$\rlap{.}^{\circ}$5 from the SMC centre. We then estimated the median abundance and the standard errors in each of the bins.  We plotted the median [Fe/H] and median radius of each concentric shell in Figure \ref{fig:feh}). To estimate the metallicity gradient, we made a linear fit using Python and the best fit is also shown in Figure \ref{fig:feh}). We obtained a negative metallicity gradient from the centre to the SMC outskirts from our sample, which is --0.062 $\pm$ 0.009 dex/deg, consistent with the previous studies suggesting that the central SMC is metal-rich. We also note that in the outer bin at $>$ 8$^\circ$ that the metallicity increases slightly and reaches e median value similar to that of the inner regions. This could be due to the lesser number of stars in the outskirts but further investigation is required to confirm it. 
\begin{equation}
    [Fe/H] = (m_{1,0} + a_1 * (b - y)_0 + a_2)/ a_3 * (b - y)_0 + a_4
    \label{equation:feh}
\end{equation}
where
\begin{equation}
    m_{1,0} = (v - b)_0 - (b - y)_0
\end{equation}

and\\
$a_1$ = -1.277 $\pm$ 0.050;  $a_2$ = 0.331 $\pm$ 0.035; $a_3$ = 0.324 $\pm$ 0.035;  $a_4$ = -0.032 $\pm$ 0.025 \\

We are also estimating the uncertainties for our [Fe/H] estimations using error propagation and it will further help as a constraint to have more significant results.


\section{Summary and Ongoing work}
In this contribution, we presented the method we used and the preliminary results from our investigation of the metallicity distribution across the SMC where we estimate [Fe/H] from \textit{Gaia} XP spectra using the synthetic Str\"{o}mgren photometry. The resulting [Fe/H] gradient of --0.062 $\pm$ 0.009 dex/deg from our study is comparable with the results of previous studies like --0.075 $\pm$ 0.011 dex/deg from \cite{2014MNRAS.442.1680D} and --0.031 $\pm$ 0.005 dex/deg from \cite{2020MNRAS.497.3746C}. Our study also validates the potential usage of \textit{Gaia} XP spectra to estimate the individual [Fe/H] values which are needed for different science cases. As we have mentioned in the previous sections, we have only explored a subset of the data available for the SMC. We also aim to apply this method to different types of stars, not only giants, within the galaxy to increase the size of our metallicity sample and reduce the uncertainties in our measurements. One of our next steps is to produce a high spatial resolution photometric map using the estimated values. Then identify the sources in different substructures and compare them with those of the main body of the SMC to shed light on their possible origins. \\
\indent We are also applying this general method we have developed in our study to stellar populations of the Large Magellanic Cloud and this can also be extended to other systems. This will provide us with the largest and most homogeneous abundance estimates of the MCs, which can be used to analyse different stellar sub-structures, metallicities of young and old stellar populations, metallicities in different regions of the MCs and so on. \\

\textbf{Acknowledgement}\\

AOO is grateful to ESA for support via the Archival Research Visitor Programme during her stay at ESTEC.

\end{document}